\documentclass[%
 aip,
cp,  
 amsmath,amssymb,
 reprint,%
]{revtex4-2}

\usepackage{graphicx}
\usepackage{dcolumn}
\usepackage{bm}
\usepackage{xcolor, ulem}

\usepackage[T1]{fontenc}
\usepackage{mathptmx} 

\begin{document}

\title{From $\bar{K}N$ Interactions to $\bar{K}$-Nuclear Quasi-Bound States}

\author{Ale\v{s} Ciepl\'{y}} 
 \email[Corresponding author: ]{cieply@ujf.cas.cz}
\author{Jaroslava Hrt\'{a}nkov\'{a}}%
\author{Ji\v{r}\'{\i} Mare\v{s}}
\affiliation{
  Nuclear Physics Institute of the Czech Academy of Sciences, 250 68 \v{R}e\v{z}, Czech Republic
}
\author{\\ Eliahu Friedman}
\author{Avraham Gal}
\affiliation{
  Racah Institute of Physics, The Hebrew University, 91904 Jerusalem, Israel
}

\author{\`{A}ngels Ramos}
\affiliation{
  Facultat de F\'{\i}sica, Universitat de Barcelona, Mart\'{\i} i Franqu\`{e}s 1, 08028 Barcelona, Spain
}

\date{\today} 

\begin{abstract}
We review the current status of our study of $K^-$-nuclear interactions and $K^-$-nuclear quasi-bound states. 
The adopted $K^-$-nuclear optical potential consists of two parts -- the single-nucleon one constructed
microscopically from chirally motivated $\bar{K}N$ amplitudes, and a phenomenological
multi-nucleon one constrained in fits to kaonic atoms data. The inclusion of multi-nucleon
absorption in our calculations of $K^{-}$ quasi-bound states in many-body systems leads to huge
widths, considerably exceeding the binding energies. If this feature is confirmed 
the observation of such states is unlikely. Finally, a development of a new microscopical model for in-medium $K^-NN$ absorption is discussed as well.
\end{abstract}

\maketitle

\section{\label{sec:into}Introduction}

The interactions of (anti)kaons with nucleons and nuclei represent an interesting topic related 
(among others) to SU(3) chiral symmetry, structure of composite systems or even to content 
of strangeness in hot and dense nuclear matter as studied in astrophysics. Here we review 
the recent progress in calculations of the characteristics of $K^-$-nuclear quasi-bound 
states using the $K^-$-nuclear optical potential constrained in fits to kaonic atoms data. 
The single-nucleon part of the potential is constructed as a coherent sum of  
$K^{-}N$ scattering amplitudes, $V^{1N}_{K^-} \sim t_{K^-N}(\rho)\, \rho$, where $\rho$ denotes 
the nuclear density and in-medium \mbox{t-matrices} are derived from the free-space ones 
by accounting for Pauli blocking. Strong energy dependence of the scattering amplitudes 
is treated self-consistently, iterating the energy at which the $K^{-}N$ (and $K^{-}$-nuclear) 
amplitude is evaluated. We note that in-medium hadron self-energies are disregarded in the present 
report as their inclusion would require a more complex treatment going beyond using the free space 
$K^{-}N$ amplitudes as input in our calculations. The self-consistent procedure leads to substantial 
downward energy shift, typically about 30 MeV or even as large as 50-100 MeV in calculations 
of $K^{-}$-atomic or $K^{-}$-nuclear states, respectively. The details of the procedure can 
be found e.g.~in \cite{Cieply:2011fy, Friedman:2016rfd, Hrtankova:2017zxw}.

It was demonstrated in \cite{Friedman:2016rfd} that the optical potential $V^{1N}_{K^-}$ does
not provide a satisfactory description of the kaonic atoms data and has to be supplemented 
with a part related to kaon interactions with two or more nucleons. In close similarity 
with earlier pion-nuclear studies, a phenomenological term $V^{mN}_{K^-}$ was considered 
in \cite{Friedman:2016rfd},
\begin{equation}
 2{\rm Re} (\omega_{K^-}) V_{K^-}^{mN} = -4 \pi\, B\, (\rho / \rho_0)^{\alpha} \, \rho \; ,
\label{eq:VmN}
\end{equation}
where $\omega_{K^-} = m_{K^-} - B_{K^-} -{\rm i}\Gamma_{K^-}/2$ stands for (a complex) kaon energy, 
$B_{K^-}$ ($\Gamma_{K^-}$) denotes the $K^-$ binding energy (width), $m_{K^-}$ is the $K^-$ mass
and $\rho_0 = 0.17$ fm$^{-3}$. The complex parameter $B$ determines the strength of the $V^{mN}_{K^-}$ 
potential that simulates processes on two nucleons or in general any {\it multi-nucleon}
contributions to the total optical potential, $V_{K^-} = V^{1N}_{K^-} + V^{mN}_{K^-}$. 

When calculating the $K^-$-atomic or $K^-$-nuclear states, the interaction of the $K^-$ meson 
with a nucleus is described by the Klein-Gordon equation
\begin{equation}
 \left[ \vec{\nabla}^2  + \tilde{\omega}_{K^-}^2 -m_{K^-}^2 
 -\Pi_{K^-}(\omega_{K^-},\rho) \right]\phi_{K^-} = 0        \;,   
\label{eq:KG}
\end{equation}
where $\tilde{\omega}_{K^-} = \omega_{K^-} - V_C$ with $V_C$ standing for the Coulomb potential.
The kaon self-energy is related directly to the $K^-$-nuclear optical potential, 
$\Pi_{K^-}(\omega_{K^-},\rho) = 2 {\rm Re}( {\omega}_{K^-}) V_{K^-}$.
Finally, we note that as the optical potential is constructed from the $\bar{K}N$ amplitudes 
evaluated at energies affected (in nuclear matter) by the kaon self-energy, the potential 
itself as well as the in-medium energy shift it causes are to be calculated self-consistently.

\section{\label{sec:KbarN}Chirally motivated $\bar{K}N$ amplitudes}

A modern treatment of low-energy meson-baryon interactions is provided by approaches 
based on chiral perturbation theory combined with coupled channel T-matrix re-summations 
techniques. The parameters of such models are fitted to low energy $K^{-}p$ total cross 
sections, the threshold branching ratios (see e.g~\cite{Feijoo:2018den for a list 
of references to the relatively old bubble-chamber experiments)} and to the strong-interaction 
characteristics of the 1s level in kaonic hydrogen measured recently by the SIDDHARTA 
collaboration \cite{Bazzi:2011zj}. Several theoretical groups presented models 
describing about equally well this set of experimental data. We refer to these approaches 
as Kyoto-Munich (KM) \cite{Ikeda:2012au}, Prague (P) \cite{Cieply:2011nq}, 
Bonn (B2, B4) \cite{Mai:2014xna}, Murcia (M$_{I}$, M$_{II}$) \cite{Guo:2012vv} 
and Barcelona (BCN) \cite{Feijoo:2018den}, with some of them providing two solutions. The first four models are compared 
in~\cite{Cieply:2016jby}. 

\begin{figure}[b]
\includegraphics[width=0.98\textwidth]{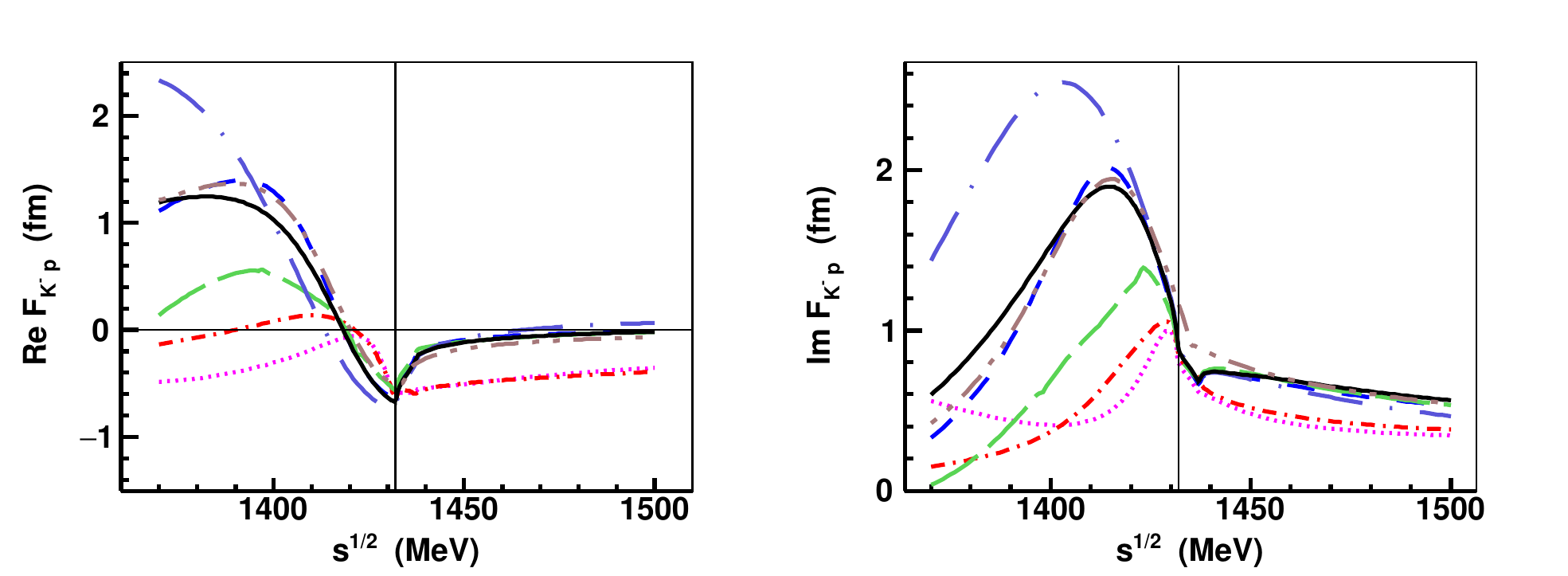} \\
\includegraphics[width=0.98\textwidth]{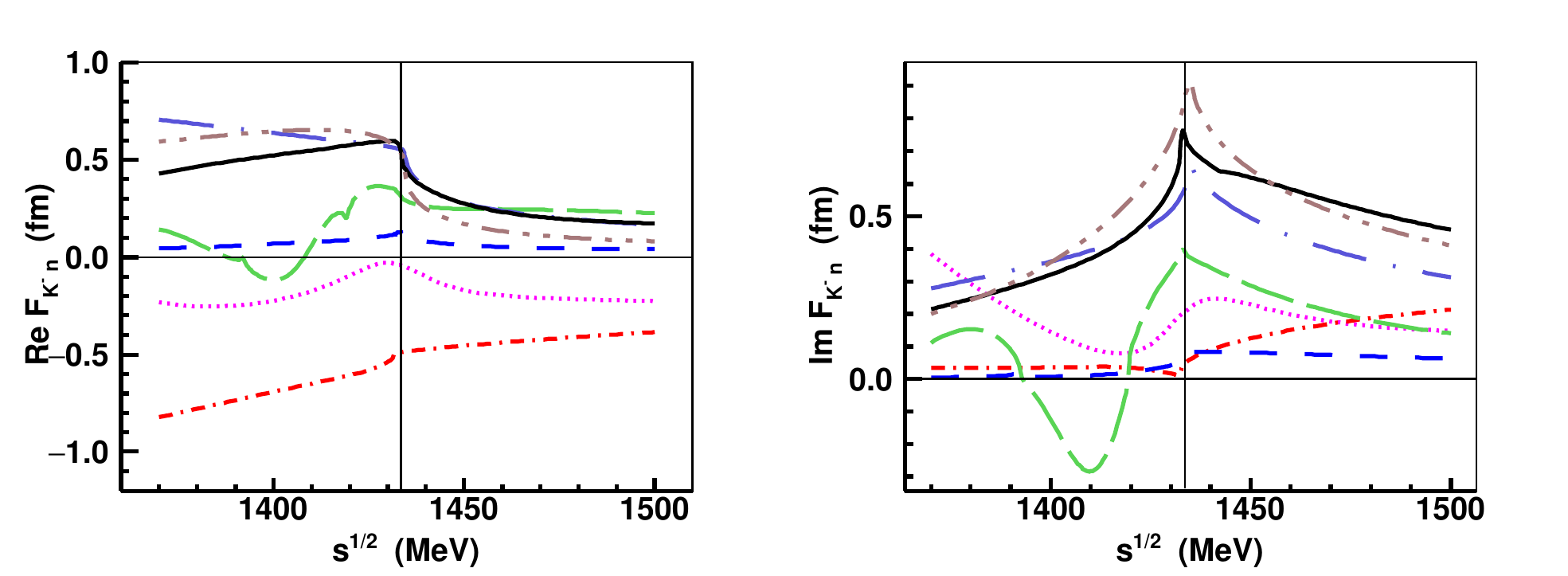} 
\caption{The $K^{-}p$ (top panels) and $K^{-}n$ (bottom panels) elastic scattering amplitudes 
generated by recent chirally motivated approaches. The various lines refer to the models:  
B$2$ (dotted, purple), B$4$ (dot-dashed, red), M$_{I}$ (dashed, blue), 
M$_{II}$ (long-dashed, green), P (dot-long-dashed, violet), BCN (dot-dot-dashed, brown), 
and KM (continuous, black). The thin vertical lines in the panels mark the pertinent 
$K^{-}N$ thresholds.
}
\label{fig:ampl}
\end{figure}

In Fig.~\ref{fig:ampl} we present the predictions of the models for $K^{-}p$ 
and $K^{-}n$ elastic amplitudes in the free space. Concerning the $K^{-}p$ amplitude, all these
{\it state-of-the-art} chiral models are in agreement in a region of energies at and above 
the $K^{-}p$ threshold. The only exception is the Bonn approach due to different treatment of off-shell effects and partial waves. The above models yield considerably different $K^{-}p$ amplitude below the threshold. On the other hand, for the $K^{-}n$ amplitude the model variations are quite large over the whole energy region. The reason is that the $I=1$ amplitudes, as well as the subthreshold $K^-p$ amplitudes, are not sufficiently restricted by the experimental data. 

In nuclear matter the free-space $K^{-}N$ amplitudes are modified due to Pauli blocking 
and hadron self-energies, the latter effectively modifying the in-medium hadron masses as well. It appears that for energies at least about $20$~MeV below the $\bar{K}N$ threshold the main 
effect comes from the Pauli blocking and can be approximated by a simple multiplication 
of the free-space $K^{-}N$ amplitudes by an energy and density dependent factor derived from considering nucleon-nucleon anti-correlations 
\cite{Waas:1996tw, Friedman:2016rfd, Hrtankova:2017zxw}.

\section{\label{sec:Katoms}Kaonic atoms}

The data base for kaonic atoms comprises of 65 experimental results for strong interaction 
level shifts and widths of so-called {\it lower states} and relative yields of transitions 
from so-called {\it upper states} to the lower states \cite{Friedman:2007zza}. This data set was used in fits performed in \cite{Friedman:2016rfd} with several versions of the $K^-$-nuclear optical potential.
First, it was found that the potential $V^{1N}_{K^-}$ constructed self-consistently from 
the in-medium $\bar{K}N$ amplitudes fails miserably when used to calculate the $K^-$-atomic
characteristics. However, good fits to the data were obtained for all models augmented with 
the phenomenological $V^{mN}_{K^-}$ term. We refer the reader to \cite{Friedman:2016rfd} and
\cite{Friedman:2019nol} for more details on the analysis and fitting procedure. Here we 
only mention that due to correlations between $\alpha$ and the complex $B$ parameter, that
define the form and strength of the $mN$ term, it was necessary to scan over the parameter
$\alpha$ when searching for the best $\chi^{2}$ value while varying the $B$ parameter. 

\begin{figure}[h]
\includegraphics[width=0.48\textwidth]{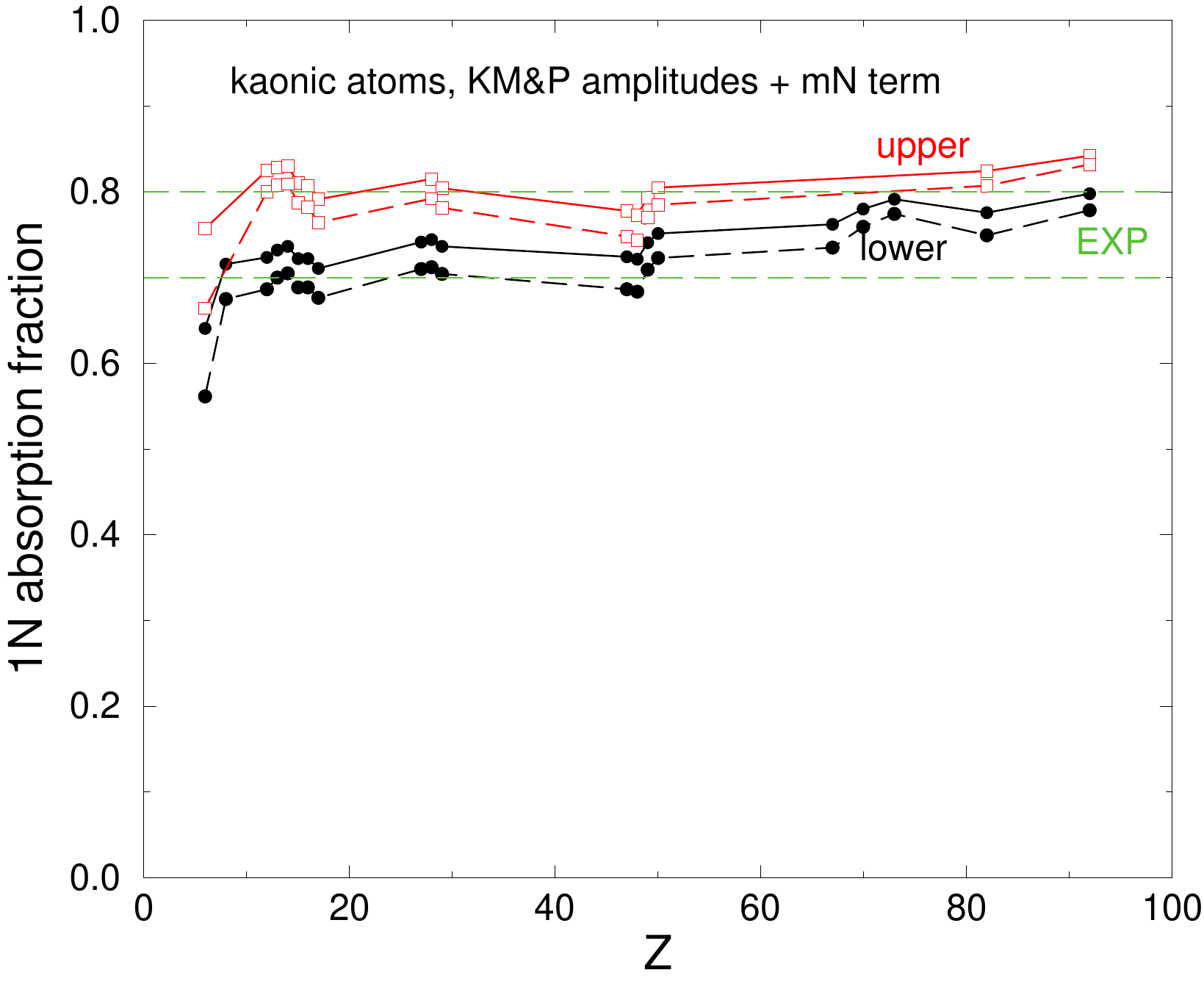} \hspace*{2mm}
\includegraphics[width=0.48\textwidth]{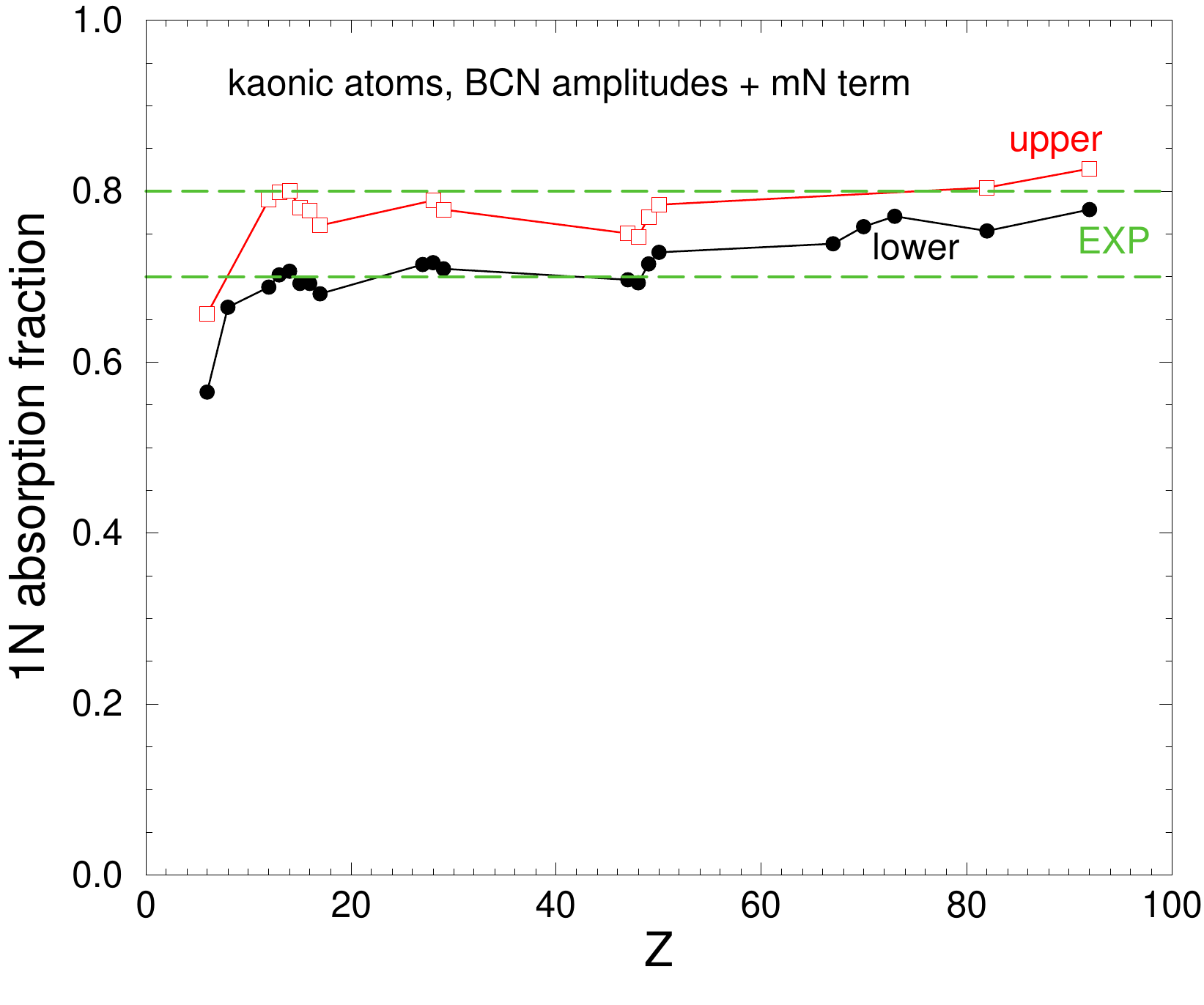} 
\caption{
Left panel: Calculated SNAF for the P and KM models, $\alpha=1$ (solid line) and $\alpha=2$ (dashed line) \cite{Friedman:2016rfd}.  Right panel: SNAF for the  BCN model.
The solid circles and open squares stand for {\it lower} and {\it upper} $K^-$-atomic states, 
respectively. }
\label{fig:snaf}
\end{figure}

In Table \ref{tab:Katom} we show the results of the fits performed with the Prague, Kyoto-Munich 
and Barcelona models. For these fits, the $\alpha$ parameter was fixed at values
very close to the best fit ones with the variations in $B$ reflecting the model uncertainties.
The achieved total $\chi^{2}$ values are very similar for all models including the Bonn and
Murcia ones. The latter ones are omitted in Table \ref{tab:Katom} as they do not satisfy 
an additional constraint of reproducing single-nucleon absorption fractions (SNAF) measured 
in bubble-chamber experiments on several nuclei from C to Br long time ago 
\cite{Davis:1968xxx, Moulder:1972up, Vandervelde:1977xxx}. A common experimental value 
SNAF$_{\rm exp} = 0.75\pm0.05$ was adopted in \cite{Friedman:2016rfd} as a further
test for the input $\bar{K}N$ models. While the B and M models were found to significantly
underestimate the SNAF value, the other three models based on the P, KM and BCN amplitudes
comply nicely with it as demonstrated in Fig.~\ref{fig:snaf}. There, the P and KM model predictions almost coincide, so we opted to show lines that represent both models in the left panel of Fig.~\ref{fig:snaf}.

\begin{table}[h]
\caption{The parameters and total $\chi^2$ values obtained in fits to kaonic atoms data. 
For the optical potentials based on the P and KM amplitudes two versions of density dependence 
were considered in the phenomenological $mN$ term, tagged as P$\alpha$ and KM$\alpha$,
respectively.}
\begin{ruledtabular}
\begin{tabular}{c|ccccc}
             &  P1          &     KM1      &      BCN     &    P2        & KM2         \\ \hline
  $\alpha$   &  1           &      1       &       1      &    2         &     2       \\ 
Re$\,B$ (fm) & -1.3$\pm$0.2 & -0.9$\pm$0.2 & -1.3$\pm$0.3 & -0.5$\pm$0.6 & 0.3$\pm$0.7 \\ 
Im$\,B$ (fm) & ~1.5$\pm$0.2 & ~1.4$\pm$0.2 & ~1.9$\pm$0.3 & ~4.6$\pm$0.7 & 3.8$\pm$0.7 \\
$\chi^2(65)$ & 125 & 123 &  129  & 125 & 123 
\end{tabular}
\end{ruledtabular}
\label{tab:Katom}
\end{table}

\section{\label{sec:Knuc}Kaonic nuclei}

The formalism outlined above was adopted to self-consistent calculations of $K^-$-nuclear quasi-bound states in selected nuclei across the periodic table~\cite{Hrtankova:2017zxw, Hrtankova:2017plb}.
The nucleus was described within a relativistic mean field model. 
It was demonstrated in the analysis of kaonic atoms 
\cite{Friedman:2016rfd} that the optical potential $V_{K^-}$ is not so well 
determined at densities larger than about 30-50\% of the nuclear density $\rho_0$. 
For this reason, two scenarios were adopted in calculations of the $K^{-}$-nuclear states. 
In the {\it full density} (FD) option the $V^{mN }_{K^-} \sim B\,(\rho/\rho_0)^\alpha\rho$ form 
was used in the entire nucleus, while in the {\it half density} (HD) option the $mN$ 
term was fixed at a constant value equal to $V_{K^-}^{mN }(0.5\rho_0)$ for densities 
$\rho \ge 0.5 \rho_0$. In addition, the amplitude Im$\,B$ in Eq.~(\ref{eq:VmN}) was multiplied
by a kinematical suppression factor to account for phase space reduction. More details can be found in~\cite{Hrtankova:2017zxw, Hrtankova:2017plb}. 

\begin{figure}[h!]
\includegraphics[width=0.48\textwidth]{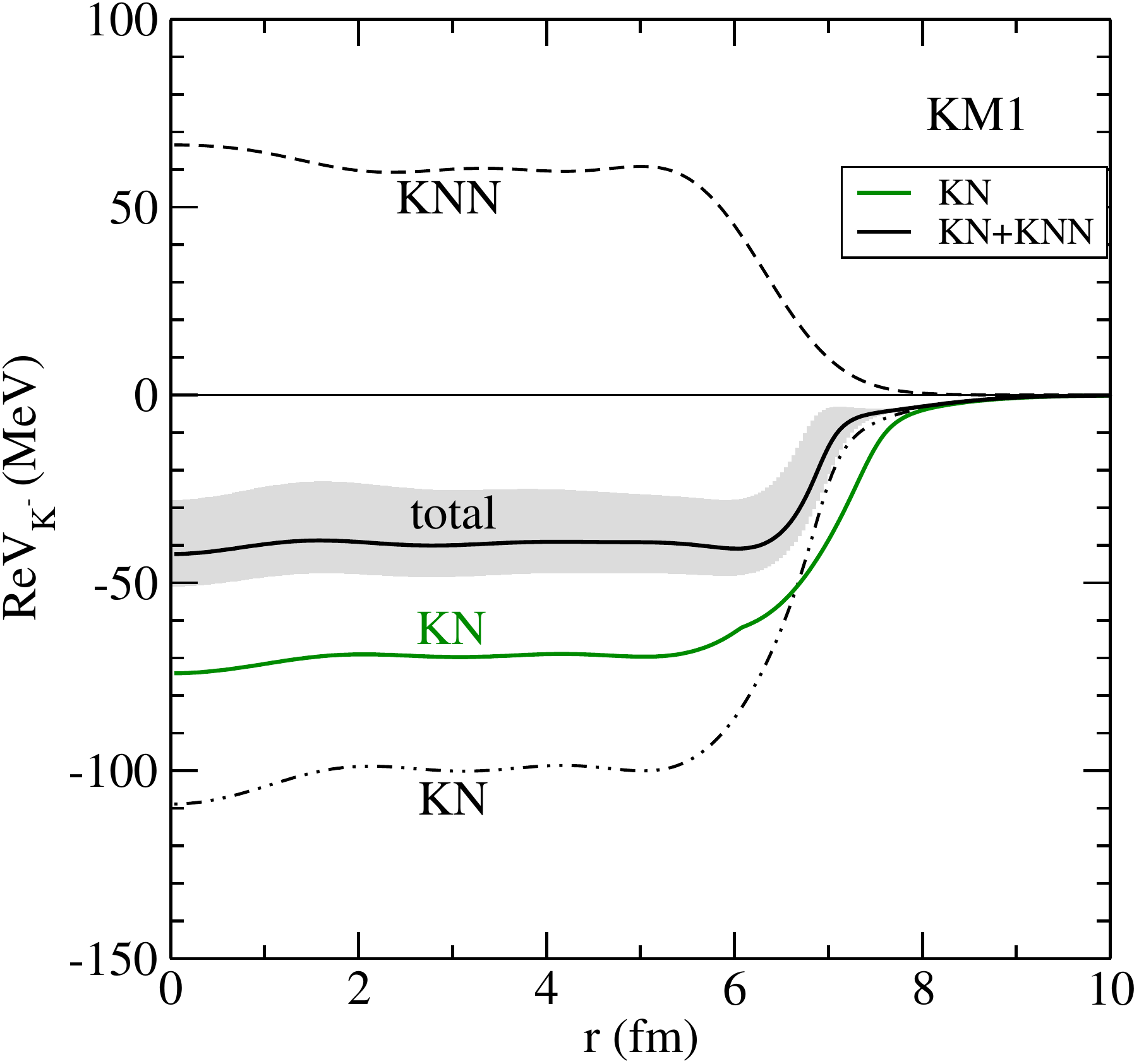} \hspace*{2mm}
\includegraphics[width=0.48\textwidth]{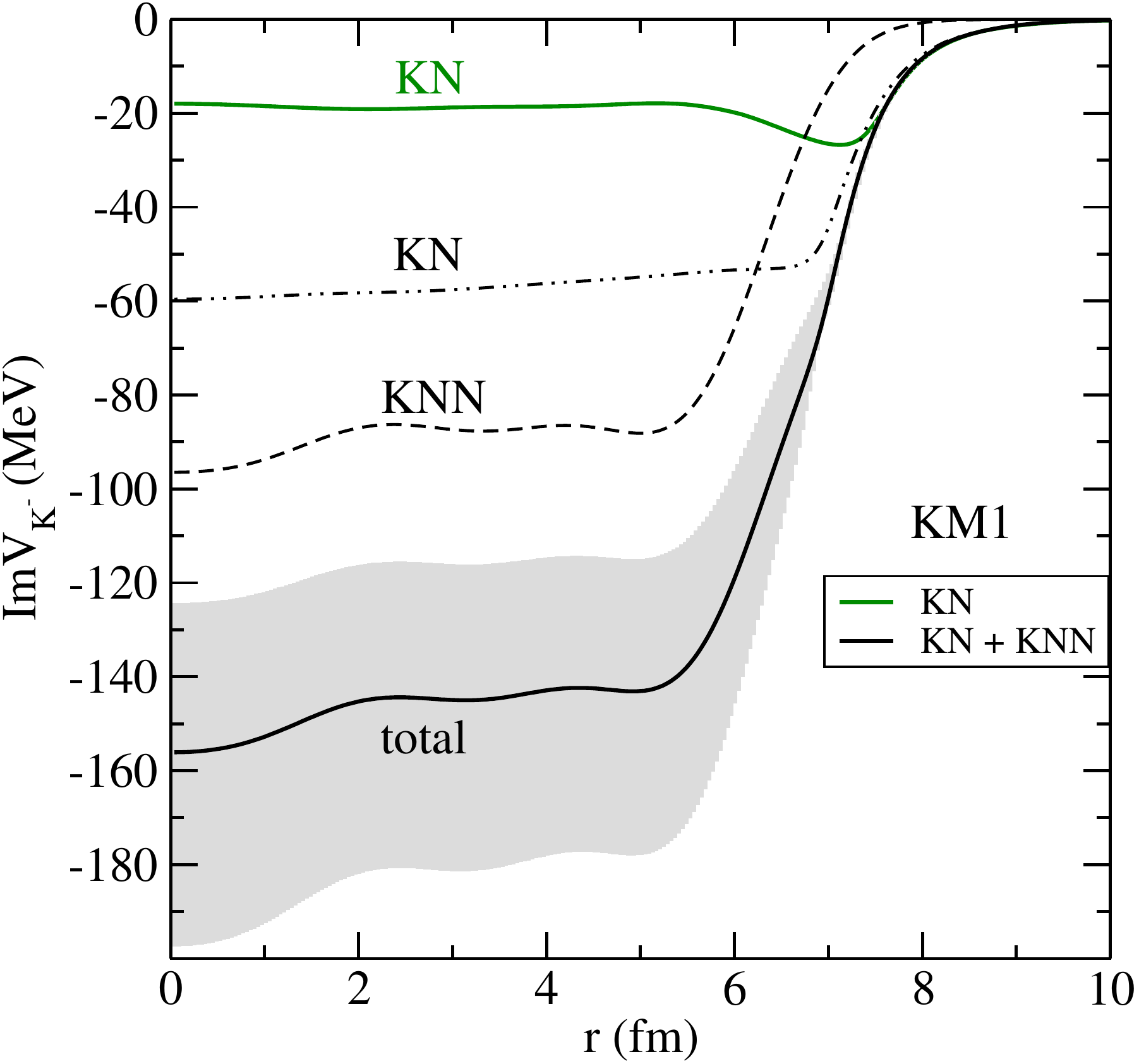} 
\caption{
The contributions from $V^{1N}_{K^-}$ and $V^{mN}_{K^-}$ potentials to the total real 
and imaginary $K^-$ optical potential in the $^{208}$Pb$+K^-$ nucleus, calculated 
self-consistently in the KM1 model and for the FD variant. 
The single-nucleon $K^-$ potential (green solid line) of the KM model 
is shown for comparison. Shaded area represent uncertainties in the phenomenological 
$mN$ part input.
}
\label{fig:Vopt}
\end{figure}

In Fig.~\ref{fig:Vopt} we compare the contributions of the $1N$ (denoted KN) and $mN$ 
(denoted KNN) parts in the KM1 version of the optical potential. Both parts are calculated 
self consistently for the $^{208}$Pb nucleus. The $1N$  component of the optical potential 
notably differs from the input $1N$ term (green solid lines) owing to different subthreshold 
energy shifts obtained upon including (or excluding) the $mN$ phenomenological term. 
The marked uncertainties reflect the variations in the $mN$ input strength parameter $B$ 
determined in kaonic atoms fit and noted in Table \ref{tab:Katom}. Whereas the resulting 
real potential depths are considerably lower than those obtained in totally phenomenological 
analyses \cite{Mares:2006vk}, the imaginary potentials are dominated by the $mN$ component 
and become extremely deep, close to 160 MeV (and much more in the KM2 optical potential). 
The dominance of the $mN$ component over the  $1N$  one is also realized in most of the other 
models based on the chirally motivated $\bar{K}N$ amplitudes employed 
in \cite{Hrtankova:2017zxw}. 

\begin{figure}
\begin{center}
\includegraphics[width=0.5\textwidth]{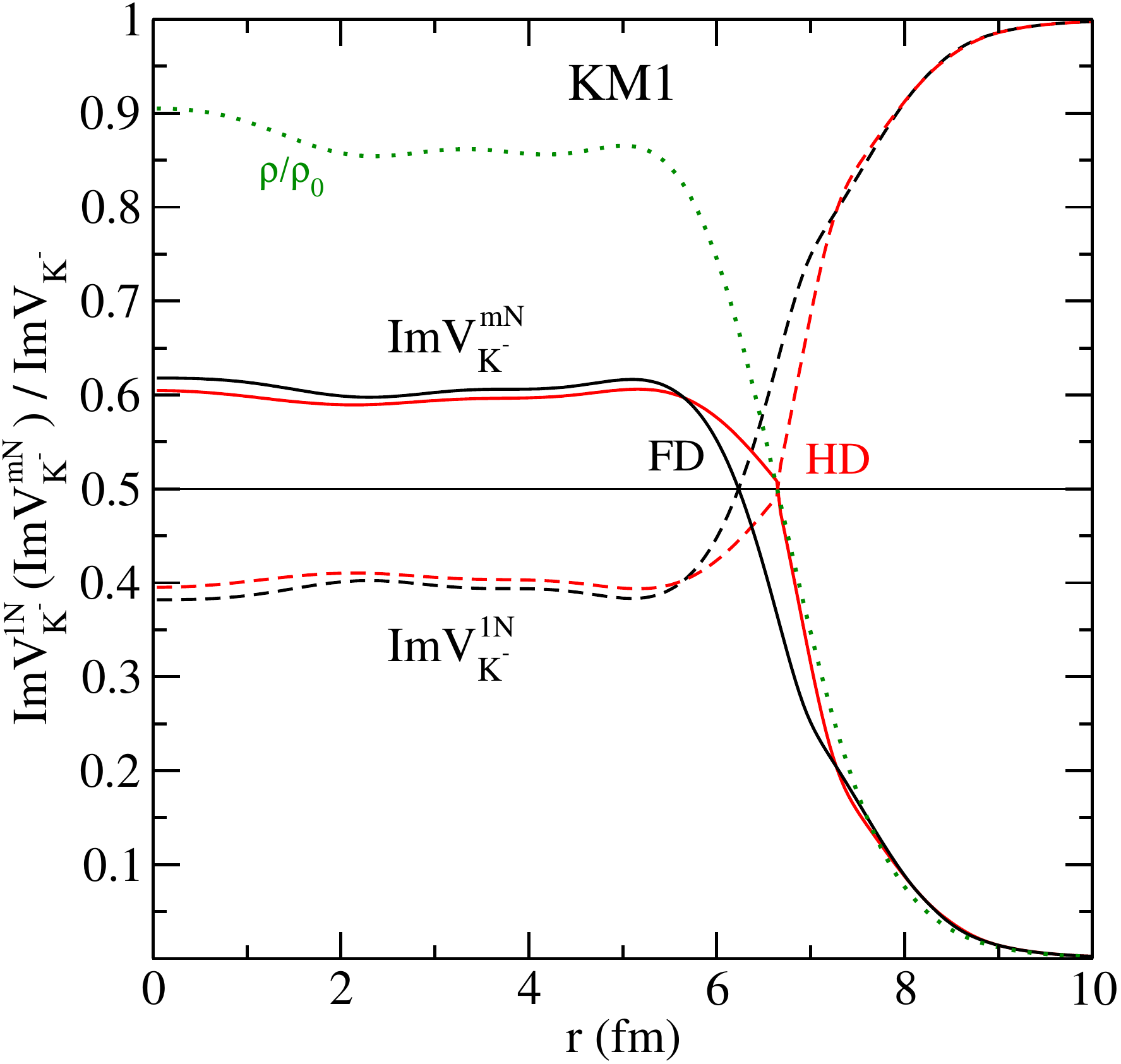}
\end{center}
\caption{
Ratios of Im$\,V_{K^-}^{1N}$ and Im$\,V_{K^-}^{mN}$ potentials to the total 
Im$\,V_{K^-}$ as a function of radius, calculated self-consistently for $^{208}$Pb$+K^-$ 
system in the KM1 model and different options for the $K^-$ multi-nucleon potential.}
\label{fig:pomer}
\end{figure}

In Fig.~\ref{fig:pomer} we present individual contributions of the $K^-$ single-nucleon (dashed
lines) and multi-nucleon (solid lines) absorption to the total $K^-$ absorption, expressed 
as a fraction of Im$\,V_{K^-}^{1N}$ and  Im$\,V_{K^-}^{mN}$ with respect to the total imaginary potential Im$\,V_{K^-}$ in $^{208}$Pb, calculated self-consistently within the KM1 model 
for the HD (red) and FD (black) options. The nuclear density distribution $\rho / \rho_0$ (thin dotted line)
is shown for comparison. Both versions of the $V_{K^-}^{mN}$ potential give very similar
fractions of single- and multi-nucleon absorption at low densities as well as in the nuclear
interior. Since the range and density dependence of the corresponding potentials are different,
the relative contribution of Im$\,V_{K^-}^{1N}$ and Im$\,V_{K^-}^{mN}$ to the $K^-$ absorption is changing with radius (density), as expected. While at the nuclear surface, the dominant
process is the $K^-$ absorption on a single nucleon, in the nuclear interior the $1N$ absorption
is reduced due to the vicinity of the $\pi\Sigma$ threshold and the $mN$ absorption prevails. 

In Table \ref{tab:Knuc1s} the binding energies $B_{K^-}$ and absorption widths $\Gamma_{K^-}$ 
are presented for the 1s levels in several $K^-$-nuclei across the periodic table, as calculated within the P and KM models. In the KN columns we also show the binding energies and widths obtained when 
only the $1N$ term is used in the optical potential. The inclusion of the $mN$ term leads 
to a drastic increase of $K^-$ widths while $K^-$ binding energies are affected only 
moderately. For most kaonic nuclei the HD option of the $V_{K^-}^{mN}$ potential yields 
$K^-$ widths of about $100$~MeV while the binding energies are much smaller than the pertinent 
widths. The situation is much worse for the FD version of the optical potential 
that even does not predict any $K^-$ bound state in the majority of nuclei. Therefore, 
our results suggest that the widths of $K^-$-nuclear quasi-bound states are considerably larger
than their binding energies, at least for nuclei with $A \ge 10$. This makes an experimental 
observation of such states very unlikely.

\begin{table}[h!]
\caption{1s $K^-$ binding energies $B_{K^-}$ and widths $\Gamma_{K^-}$ (in MeV) in various nuclei calculated within the KM and P models with the
$V_{K^-}^{1N}$ potential (denoted KN); plus a phenomenological $V_{K^-}^{mN}$ term 
for the HD$\alpha$ and FD$\alpha$ options (see text for details).
}
\begin{ruledtabular}
\begin{tabular}{rc|ccccc|ccccc}
 &  & \multicolumn{5}{c}{KM model}  & \multicolumn{5}{c}{P model} \\ \hline
 &  &  KN &  HD1 & FD1 &  HD2 & FD2 &  KN &  HD1 & FD1 &  HD2 & FD2 \\ \hline 
 $^{16}$O  & $B_{K^-}$ & 45 & $\; 34$ &  not & $\; 48$ & not  & 64 & 49 & not & $\; 63$ & not  \\ 
      & $\Gamma_{K^-}$ & 40 & 109 &  bound & 121  & bound & 25 & 94 & bound & 117  & bound \\ \hline
 $^{40}$Ca  & $B_{K^-}$ & 59 & $\; 50$ & not & $\; 64$ & not  & 81 & 67 & not & $\; 82$ & not  \\   
      & $\Gamma_{K^-}$ & 37 & 113 & bound & 126  & bound & 14 & 95 & bound & 120  & bound \\ \hline
 $^{208}$Pb  & $B_{K^-}$ & 78 & $\; 64$ & $\; 33$ & $\; 80$ & 
$\; 53$  & 99 & 82 & $\; 36$ & $\; 96$ & $\; 47$  \\ 
      & $\Gamma_{K^-}$ & 38 & 108 & 273 & 122  & 429 & 14 & 92 & 302 & 117  & 412 \\
\end{tabular}
\end{ruledtabular}
\label{tab:Knuc1s}
\end{table}

\section{\label{sec:KbarNN}$\bar{K}NN$ in-medium absorption}

The introduction of the phenomenological $V^{mN}_{K^-}$ potential was essential for a successful 
description of $K^{-}$-atomic data. However, one may argue that an energy-dependent two-nucleon 
potential of a different form could do the job as well, and may not lead to such a drastic 
increase of the predicted decay widths of the $K^{-}$-nuclear states. In this Section we briefly 
report on a development of a microscopical model describing the $K^-$ absorption on two nucleons 
in nuclear matter. The formalism follows closely the approach adopted in \cite{Nagahiro:2011fi} 
to derive the $\eta'NN$ optical potential. The $K^{-}$ absorption on two nucleons is described 
within a meson-exchange picture with the $K^{-}NN$ self-energy modeled using chirally motivated
$\bar{K}N$ amplitudes modified due to Pauli blocking. In actual calculations, 
the P and BCN models were used providing a comparison of the results and a measure 
of the sensitivity to a particular $\bar{K}N$ model.

\begin{figure}[h!]
\includegraphics[width=0.48\textwidth]{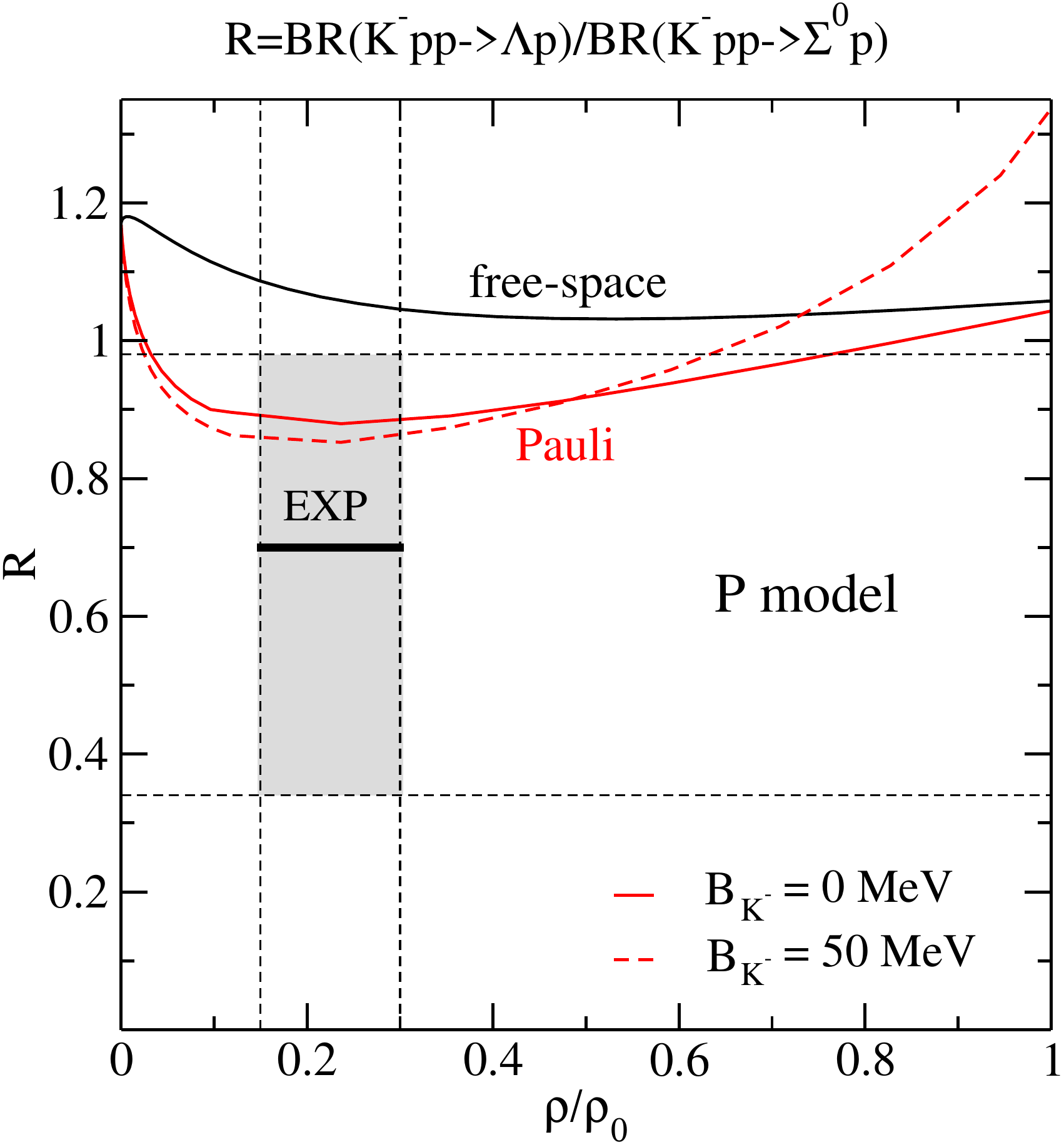} \hspace*{2mm}
\includegraphics[width=0.48\textwidth]{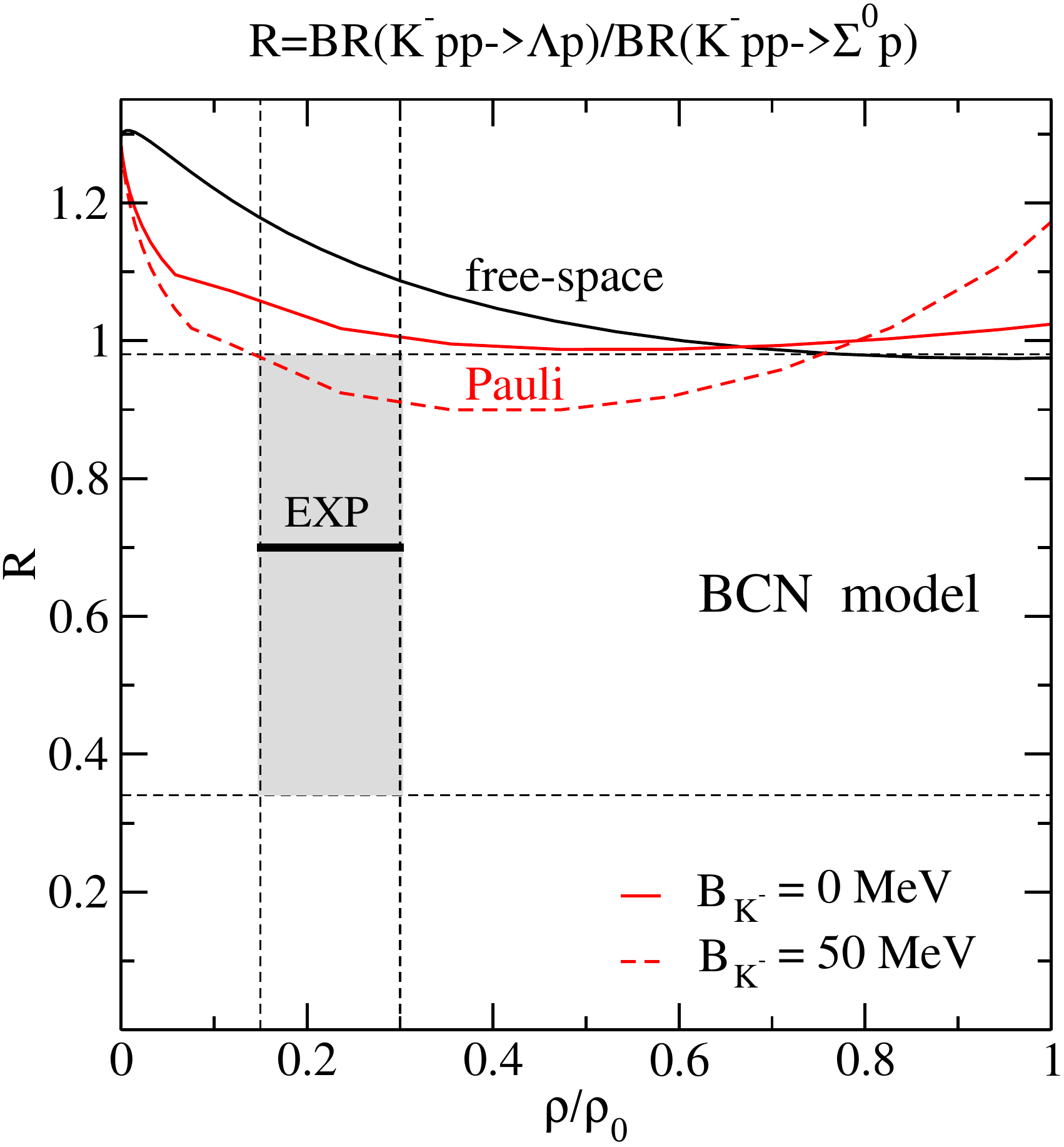} 
\caption{
The ratio of branching ratios for the in-medium $K^-pp\rightarrow \Lambda p$ and 
$K^-pp\rightarrow \Sigma^0 p$ reactions. The gray rectangle marks an estimated region 
of densities probed by low-energy antikaons and the experimental error band. 
Left - P model, right - BCN model.
}
\label{fig:BR}
\end{figure}

We do not dwell on technical details of the $\bar{K}NN$ model here as they are prepared 
for publication elsewhere \cite{Hrtankova:2019jky}. Instead, we promote the work by picking 
one result to demonstrate the applicability of the model. It relates to a recent AMADEUS 
measurement of the $\Lambda p$ to $\Sigma^0 p$ production rate in $K^-NN$ quasi-free absorption 
reported in \cite{DelGrande:2018sbv},
\begin{equation}
     R \;=\; 
    \frac{{\rm BR}(K^-pp\rightarrow \Lambda p)}{{\rm BR}(K^-pp\rightarrow \Sigma^0 p)} 
    \;=\; 0.7 \pm 0.2(stat.)^{+0.2}_{-0.3}(syst.) \; .
\end{equation}
In Fig.~\ref{fig:BR} taken from \cite{Hrtankova:2019jky} we show the ratio predicted
with the new $\bar{K}NN$ model when either the free space (black lines) or Pauli blocked 
(red lines) $\bar{K}N$ amplitudes are employed. The density dependence of the rate 
is presented using the P and BCN amplitudes for two settings of the $\bar{K}NN$ model,
assuming the $K^{-}$ binding energy in nuclear matter to be either $B_{K^-} = 0$ or $B_{K^-} = 50~\rho/\rho_0$ MeV.
In the experimentally relevant region of densities, the ratio $R$ calculated 
with Pauli blocked amplitudes turns out smaller and in better agreement with 
the measured value. This feature manifests the relevance of medium modifications 
of the chirally motivated amplitudes.

\section{\label{sec:res}Brief summary}

\begin{itemize}
\item The up-to-date (NLO) chirally motivated $\bar{K}N$ models provide very different
predictions for the $K^{-}N$ amplitudes at subthreshold energies. As in-medium kaons probe
energies as far as 50-100 MeV below the $\bar{K}N$ threshold a realistic treatment 
of the energy dependence including Pauli blocking is essential.
\item 
$K^-$ optical potentials derived from chirally-inspired $K^-N$ interaction models need to be supplemented by an additional phenomenological density-dependent term representing $K^-$
multi-nucleon interactions in order to get satisfactory global fit to kaonic-atom strong-interaction data.
The P, KM and BCN models are favored satisfying the additional constraint of $1N$ to $mN$ absorption rate. 
\item The inclusion of multi-nucleon absorption in the calculations of $K^{-}$ quasi-bound
states in many-body systems leads to huge widths, considerably exceeding the binding energies. 
If this feature is confirmed the observation of such states is unlikely. The conclusion 
does not necessarily apply to few body $K^-$-nucleons systems.
\item A microscopical model for $K^-NN$ absorption in nuclear matter has been developed using 
chirally motivated $\bar{K}N$ amplitudes. The results look encouraging, 
the ratio of $\Lambda p$ to $\Sigma^0 p$ production measured by AMADEUS is reproduced 
by the model when the in-medium amplitudes are employed.
\end{itemize}

\begin{acknowledgments}
This work was supported by the Czech Science Foundation GACR grant 19-19640S.
\end{acknowledgments}

\nocite{*}
\bibliography{MENUcit}

\end{document}